\documentstyle[neuchatel]{article}
\newcommand{\be}{\begin{equation}}
\newcommand{\ee}{\end{equation}}
\begin {document}
\def\titleline{
%%%%%%%%%%%%%%%%%%%%%%%%%%%%%%%%%%%%%%%%%%%%%%
%                                            %
% Insert now the text of your title.         %
% Make a linebreak in the title with         %
%                                            %
%            \newtitleline                   %
%                                            %
%%%%%%%%%%%%%%%%%%%%%%%%%%%%%%%%%%%%%%%%%%%%%%
BPS-spectra in four dimensions from $M$-theory
%%%%%%%%%%%%%%%%%%%%%%%%%%%%%%%%%%%%%%%%%%%%%%
}
\def\authors{
%%%%%%%%%%%%%%%%%%%%%%%%%%%%%%%%%%%%%%%%%%%%%%
%                                            %
%  Insert now the name (names) of the author %
%  (authors).                                %
%  In the case of several authors with       %
%  different addresses use e.g.              %
%                                            %
%             \1ad  , \2ad  etc.             %
%                                            %
%  to indicate that a given author has the   %
%  address number 1 , 2 , etc.               %
%                                            %
%%%%%%%%%%%%%%%%%%%%%%%%%%%%%%%%%%%%%%%%%%%%%%
M{\aa}ns Henningson
%%%%%%%%%%%%%%%%%%%%%%%%%%%%%%%%%%%%%%%%%%%%%
}
\def\addresses{
%%%%%%%%%%%%%%%%%%%%%%%%%%%%%%%%%%%%%%%%%%%%%%
%                                            %
% List now the address. In the case of       %
% several addresses list them by numbers     %
% using e.g.                                 %
%                                            %
%             \1ad , \2ad   etc.             %
%                                            %
% to numerate address 1 , 2 , etc.           %
%                                            %
%%%%%%%%%%%%%%%%%%%%%%%%%%%%%%%%%%%%%%%%%%%%%%
Theory Division, CERN\\
CH-1211 Geneva 23, Switzerland\\
{\tt henning@nxth04.cern.ch}
%%%%%%%%%%%%%%%%%%%%%%%%%%%%%%%%%%%%%%%%%%%%%%%
}
\def\abstracttext{
%%%%%%%%%%%%%%%%%%%%%%%%%%%%%%%%%%%%%%%%%%%%%%%
%                                             %
% Insert now the text of your abstract.       %
%                                             %
%%%%%%%%%%%%%%%%%%%%%%%%%%%%%%%%%%%%%%%%%%%%%%%
I review my work together with Piljin Yi \cite{Henningson-Yi} on the
spectrum of BPS-saturated states in $N = 2$ supersymmetric Yang-Mills
theories. In an $M$-theory description, such states are realized as
certain two-brane configurations. We first show how the central
charge of the $N = 2$ algebra arises from the two-form central charge
of the eleven-dimensional supersymmetry algebra, and derive the
condition for a two-brane configuration to be BPS-saturated. We then
discuss how the topology of the two-brane determines the type of
BPS-multiplet. Finally, we discuss the phenomenon of marginal
stability and show how it is related to the mutual non-locality of
states.
%%%%%%%%%%%%%%%%%%%%%%%%%%%%%%%%%%%%%%%%%%%%%%%%%%%%%%%%%%%%%%%
}
\makefront
%%%%%%%%%%%%%%%%%%%%%%%%%%%%%%%%%%%%%%%%%%%%%%%%
%                                              %
%  Insert now the remaining parts of           %
%  your article.                               %
%                                              %
%%%%%%%%%%%%%%%%%%%%%%%%%%%%%%%%%%%%%%%%%%%%%%%%
\footnotetext[1]{Talk presented at the workshop `Quantum aspects of
gauge theories, supersymmetry and unification, Neuch\^atel
University, 18-23 September 1997, to appear in the proceedings. \\
{\tt hep-th/9711076} \\ CERN-TH/97-319 \\ November 1997}
\section{Introduction}
A recurring theme in many of the developments in string theory and
supersymmetric field theory over the last few years has been the
study of the spectrum of BPS-saturated states. The best known example
is probably theories with $N = 2$ extended supersymmetry in
four-dimensional Minkowski space ${\bf R}^{1, 3}$, where the
supertranslations algebra reads
\be
\{ \eta_{\alpha a}, \bar{\eta}_{\beta b} \} = \delta_{ab}
(\gamma^\mu)_{\alpha \beta} P_\mu + \epsilon_{a b}
\left(\Pi^+_{\alpha \beta} Z + \Pi^-_{\alpha \beta} \bar{Z} \right) .
\label{supertranslations}
\ee
The supercharges $\eta_{\alpha a}$, $a = 1, 2$ are Majorana spinors
and transform as a doublet under the $SU(2)_R$ symmetry of the $N =
2$ algebra. $P_\mu$ is the momentum and $Z$ is the complex central
charge. The projection operators on positive and negative chirality
are denoted $\Pi^\pm_{\alpha \beta}$.

The central charge $Z$ is in general a linear combination of
conserved Abelian charges, such as for example electric and magnetic
charges $n_e$ and $n_m$ and quark number charges $S$, i.e.
\be
Z = a \cdot n_e + a_D \cdot n_m + m \cdot S .
\ee
The dependence of the coefficients $a$, $a_D$ and $m$ on the
parameters and the moduli of the theory can often be described by
Seiberg-Witten theory \cite{Seiberg-Witten}. This amounts to
introducing an auxiliary Riemann surface $\Sigma$ on which a closed
meromorphic differential $\lambda$ is defined. To a set of quantum
numbers $n_e$, $n_m$ and $S$ corresponds a homology class $[C]$ on
$\Sigma$, and the central charge in this sector is then given by the
integral of $\lambda$ along a representative $C$ of $[C]$:
\be
Z = \int_C \lambda . \label{SWformula}
\ee
Furthermore, the intersection number $[C] \cdot [C^\prime]$ of two
homology classes $[C]$ and $[C^\prime]$ has an interesting
interpretation. It is given by
\be
[C] \cdot [C^\prime] = n_e n_m^\prime - n_m n_e^\prime .
\ee
In a first-quantized treatment of a particle with charges
corresponding to the homology class $[C^\prime]$ in a background
containing a particle with charges corresponding to the homology
class $[C]$, the `wave-function' is really a section of the line
bundle $L$ over ${\bf R}^3 - \{ 0 \}$, whose Chern number $c_1 (L)$
equals $[C] \cdot [C^\prime]$. A non-zero intersection number thus
means that $L$ is topologically non-trivial. The physical
interpretation is that these particles then are mutually non-local,
such as for example an electron and a magnetic monopole.

The mass $M$ of a unitary representation of the $N = 2$ algebra must
obey the BPS-bound
\be
M \geq | Z | .
\ee
Representations that saturate this bound are of particular interest
\cite{Witten-Olive}. If we limit ourselves to ${\rm spin} \leq 1$,
they are either vectormultiplets (i.e. a gauge boson and its
superpartners) or hypermultiplets (i.e. a quark and its
superpartners). The BPS-saturated states enjoy an important stability
property: Generically, such a state cannot decay, simply because it
has the minimal mass in its charge sector. The exception is at a
domain wall of marginal stability in the moduli space of vacua, where
the phases of the central charges of three BPS-saturated states are
equal. It might then be possible for the heaviest particle to decay
into the two lighter ones (provided of course that the quantum
numbers are conserved). The heaviest particle would then be absent
from the spectrum on the other side of the domain wall of marginal
stability.

To get a simple example of marginal stability, we can consider the
case of a hypermultiplet quark in the background of a hypermultiplet
dyon at weak coupling. The fermionic component of the hypermultiplet
may have normalizable modes, each of which gives rise to a double
degeneracy of dyonic states
\cite{Jackiw-Rebbi}. If the normalizable fermion mode disappears
somewhere in the moduli space, so does the extra dyonic state. The
interpretation is that it has experienced a marginal decay into the
other dyonic state and a quark. The number of normalizable fermion
modes is given by an index theorem, and in particular one can
calculate the number of such modes that appear or disappear as the
domain wall of marginal stability is crossed
\cite{Henningson}. The result equals the Chern number of the
`wave-function' line bundle for the quark in the dyon background that
we discussed above. Thus, at least in this case, a BPS-saturated
particle can only experience marginal decay into two other particles
if these are mutually non-local.

We would like to get a more general understanding of exactly when the
phenomenon of marginal stability really does occur, or, more
generally, what is the spectrum of BPS-saturated states at a given
point in the moduli-space? We will address this problem in the
context of an $M$-theory description of these models, following
\cite{Witten}. (Other approaches are described in
\cite{Bilal-Ferrari}\cite{Klemm-Lerche-Mayr-Vafa-Warner}.)

\section{The $M$-theory realization}
We consider $M$-theory on an eleven-manifold which is a direct
product of four-dimensional Minkowski space ${\bf R}^{1, 3}$ and some
seven manifold $X^7$. We also include a five-brane, the world-volume
of which fills ${\bf R}^{1, 3}$ and defines a two-dimensional compact
surface $\Sigma$ in $X^7$ so that space-time Poincar\'e invariance is
unbroken. Our four-dimensional field-theory will then be the infrared
limit of the world-volume theory on the five-brane.

Unbroken supersymmetries are generated by spinor fields that

i) are covariantly constant with respect to the background metric.

ii) have positive chirality with respect to the tangent space of each
five-brane world-volume element.

\noindent
To get an $N = 2$ theory in ${\bf R}^{1, 3}$, we must thus impose
certain restrictions on the background metric and the five-brane
configuration: First of all, we will take $X^7$ to be a manifold of
$SU(2)$ holonomy. In fact, this means that $X^7$ is a direct product
of ${\bf R}^3$ and a four-manifold $Q^4$ of $SU(2)$ holonomy, i.e.
the eleven-manifold on which $M$-theory is defined is of the form
\be
M^{1, 10} \simeq {\bf R}^{1, 3} \times {\bf R}^3 \times Q^4 ,
\ee
(The double cover of) the Lorentz group of ${\bf R}^3$ will then
constitute the $SU(2)_R$ symmetry of the $N = 2$ algebra. The $SU(2)$
holonomy of $Q^4$ means that this four-manifold is a hyper-K\"ahler
manifold, i.e. it admits a two-sphere $S^2$ of inequivalent complex
structures $J$. Equivalently, $Q^4$ can be described as a Ricci-flat
K\"ahler manifold, and therefore admits a covariantly constant
holomorphic two-form $\Omega$. The relationship between these two
descriptions is as follows:  Given a choice of complex structure $J$,
i.e. a point on the two-sphere $S^2$, we have $\Omega \sim K^\prime +
i K^{\prime \prime}$, where $K^\prime$ and $K^{\prime \prime}$ are
the K\"ahler forms corresponding to two other complex structures
$J^\prime$ and $J^{\prime \prime}$ such that $J$, $J^\prime$ and
$J^{\prime \prime}$ are all orthogonal.

To leave the $SU(2)_R$ symmetry unbroken, the five-brane world-volume
must lie at a single point $p$ in ${\bf R}^3$ and define some
two-dimensional surface $\Sigma$ in $Q^4$. To preserve $N = 2$
supersymmetry in ${\bf R}^{1, 3}$, $\Sigma$ must be a supersymmetric
surface, which in the case under consideration means that it must be
holomorphically embedded with respect to some complex structure $J$
on $Q^4$. The unbroken supersymmetry generators are then of the form
\be
\eta_{\alpha a} = \Pi^+_{\alpha \beta} \zeta^+_i Q_{\beta a i} +
\Pi^-_{\alpha \beta} \epsilon_{a b} \zeta^-_i Q_{\beta b i} .
\label{susygenerators}
\ee
Here $Q_{\beta b i}$ are the supercharges of $M$-theory with the
11-dimensional spinor index decomposed into spinor indices for $SO(1,
3)$, $SO(3)$ and $SO(4)$. $\zeta^\pm_i$ are the covariantly constant
spinors on $Q^4$ whose existence follows from the $SU(2)$ holonomy.
They have been chosen to have definite and opposite chirality with
respect to the tangent space of each element of $\Sigma$. This is
possible because $\Sigma$ is holomorphically embedded with respect to
the complex structure $J$ on $Q^4$. The $\eta_{\alpha a}$ thus have
positive chirality with respect to the five-brane world-volume. (The
$\Pi^\pm_{\alpha \beta}$ project on definite and opposite chirality
in ${\bf R}^{1, 3}$.) They are Majorana spinors in ${\bf R}^{1, 3}$
because the $Q_{\beta b i}$ are Majorana spinors in $M^{1, 10}$ and
$\zeta^+_i$ and $\zeta^-_i$ are each others charge conjugates on
$Q^4$. ($\epsilon_{a b}$ is the charge conjugation matrix on ${\bf
R}^3$.)

We have described a particular vacuum state of an $N = 2$ theory in
${\bf R}^{1, 3}$ in terms of a background metric and a five-brane
configuration. We now wish to describe excitations around this vacuum
in terms of a two-brane, the world-volume of which traces a
world-line in ${\bf R}^{1, 3}$, lies at a single point $p$ in ${\bf
R}^3$ and defines a two-dimensional surface $D$ in $Q^4$. The mass
$M$ of such a two-brane is proportional to the area of $D$, so to get
a state of finite mass, we will consider an open surface $D$, whose
boundary $C = \partial D$ lies on $\Sigma$. This means that the
two-brane ends on the five-brane
\cite{Strominger}. The quantum numbers of the corresponding state are
then given by the homology class $[C]$ of $C$ on $\Sigma$.

It now follows from the 11-dimensional supertranslations algebra in
the presence of a two-brane
\cite{Townsend}
\be
\{ Q_A, \bar{Q}_B \} = (\Gamma^M)_{A B} P_M + (\Gamma^{M N})_{A B}
Z_{M N} ,
\ee
where the two-form central charge $Z_{M N}$ is given by
\be
Z^{M N} \sim \int_D d X^M \wedge d X^N
\ee
that the supercharges (\ref{susygenerators}) fulfill the $N = 2$
algebra (\ref{supertranslations}) in ${\bf R}^{1, 3}$ with the
central charge $Z$ given by
\be
Z = \int_D \Omega_D .
\ee
Here $\Omega_D$ is the pullback to $D$ of the two-form $\Omega =
\bar{\zeta}^- \gamma_{s t} \zeta^+ d X^s \wedge d X^t$. As the
notation suggests, $\Omega$ is in fact the covariantly constant
holomorphic two-form on $Q^4$. Provided that $\Omega$ is exact in a
neighborhood of $D$, i.e. that $\Omega = d \lambda$ for some one-form
$\lambda$, we can use Stokes' theorem to recover the Seiberg-Witten
expression (\ref{SWformula}) for the central charge as the integral
of $\lambda$ along $C$.

\section{The BPS-saturated spectrum}
We have seen that the mass $M$ and the central charge $Z$ of the
two-brane are given by the area of the surface $D$ and the integral
over $D$ of the holomorphic two-form $\Omega$ respectively. A short
calculation shows that the BPS-bound for the mass is saturated if and
only if the pullback of the K\"ahler form $K$ to $D$ vanishes
identically and the phase of the Hodge dual (with respect to the
induced metric on $D$) of $\Omega$ is a constant. The second
condition in fact means that there is a second K\"ahler form
$K^{\prime \prime}$ whose pullback to $D$ vanishes. This implies that
the surface $D$ is holomorphically embedded with respect to the
complex structure $J^\prime$ which is orthogonal to the complex
structures $J$ and $J^{\prime \prime}$ corresponding to $K$ and
$K^{\prime \prime}$. Note that given the complex structure $J$ with
respect to which the surface $\Sigma$ describing the five-brane is
holomorphically embedded, there is a circle $S^1$ of orthogonal
complex structures $J^\prime$ corresponding to the phase of the
central charge $Z$.

Given a set of quantum numbers $n_e$, $n_m$ and $S$, i.e. a homology
class $[C]$ on $\Sigma$, we can calculate the corresponding central
charge $Z$. Its phase determines a complex structure $J^\prime$ on
$Q^4$, and a BPS-state with these quantum numbers corresponds to a
surface $D$ which is holomorphically embedded with respect to
$J^\prime$ and whose boundary $C = \partial D$ lies on $\Sigma$ and
represents the homology class $[C]$. The problem of determining the
BPS-saturated spectrum is thus equivalent to the problem of
determining for which homology classes $[C]$ such a surface $D$
exists. The construction of such a $D$ is facilitated by the fact that
its bulk behavior is completely determined by its boundary $C$ by
analytic continuation. On the other hand one should note, however,
that a generic surface $D$ only intersects the surface $\Sigma$ at
isolated points, and not along a curve. Also, a $D$ which does
intersect $\Sigma$ along a closed curve $C$ might run off to infinity
somewhere in the bulk. Given $[C]$, it is therefore in general a
non-trivial problem to establish the existence of a corresponding
surface $D$.

In certain cases, and for special values of the moduli, it is
possible to find exact solutions for such surfaces $D$. One can also
do numerical studies. Comparison with known results in $N = 2$ super
Yang-Mills theory led us to formulate the following conjecture (see
also \cite{Mikhailov}):

{\it BPS-saturated hypermultiplets and vectormultiplets correspond to
surfaces $D$ with the topology of a disc or a cylinder respectively.}

\noindent
Surfaces of more complicated topology do not seem to arise for
generic values of the moduli, corresponding to the absence of other
BPS-saturated multiplets of ${\rm spin} \leq 1$.

Since a BPS-saturated state cannot decay at a generic point in moduli
space, it must in general be possible to accommodate an infinitesimal
deformation of the surface $\Sigma$ describing the five-brane by an
infinitesimal deformation of the surface $D$ describing the two-brane
so that it is still holomorphically embedded and ends on $\Sigma$.
However, a BPS-saturated state can decay into two other BPS-saturated
states of appropriate quantum numbers at a domain wall of marginal
stability where the three central charges have the same phase. This
must mean that the surface $D$ of the heaviest particle degenerates
into two surfaces touching in isolated points and does not exist on
the other side of the domain wall of marginal stability.

In this way, a hypermultiplet can decay into two other
hypermultiplets when the corresponding disc degenerates into two
discs touching in a point. The intersection number $[C] \cdot
[C^\prime]$ of the corresponding homology classes thus equals $\pm
1$. For example, a quark-dyon bound state in $SU(2)$ Yang-Mills
theory with fundamental flavors can decay into a dyon with $(n_e,
n_m) = (2 n, 1)$ and a quark with $(n_e^\prime, n_m^\prime) = (1, 0)$
so that indeed $n_m n_e^\prime -  n_e n_m^\prime = 1$. Similarly, a
vectormultiplet can decay into two hypermultiplets when the
corresponding cylinder degenerates into two discs touching in two
points, i.e. the intersection number must equal $\pm 2$. For example,
a $W$-boson in $SU(2)$ Yang-Mills theory can decay into two dyons
with $(n_e, n_m) = (2 n, 1)$ and $(n_e^\prime, n_m^\prime) = (2 - 2
n, -1)$ so that indeed $n_m n_e^\prime -  n_e n_m^\prime = 2$. We see
that in both cases, the mutual non-locality of the constituent states
is a necessary (and possibly sufficient) requirement for the decay to
take place.

\end{document}